\PassOptionsToPackage{dvipsnames,table}{xcolor} 

\documentclass{vgtc}                          

\graphicspath{{figures/}{pictures/}{images/}{./}} 

\usepackage{times}                     

\usepackage{eso-pic}

\usepackage{listings}

\lstdefinelanguage{CSharp}{
	morekeywords={public,private,class,static,void,int,float,double,bool,string,if,else,return,new,true,false},
	sensitive=true,
	morecomment=[l]{//},
	morecomment=[s]{/*}{*/},
	morestring=[b]",
}

\usepackage{mathptmx}                  

 \onlineid{1262}

\vgtccategory{Research}

\title{Dichoptic Opacity: Managing Occlusion in Stereoscopic Displays via Dichoptic Presentation}

\author{George Bell\thanks{e-mail: g.bell1@newcastle.ac.uk} %
\and Alma Cantu}
\affiliation{\scriptsize Newcastle University}

\teaser{
    \centering
    \includegraphics[alt={An image containing two pairs of transparent cubes with water textures and containing pyramids visible within each. The left pair show a consistent transparency whereas the right shows an inconsistent transparency with the pyramid more visible in one than the other},width=0.95\linewidth]{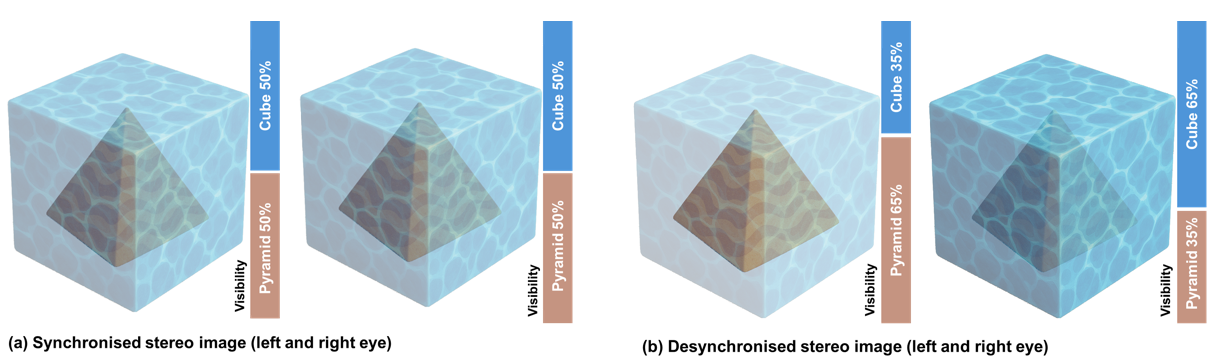}
    \caption{Non-dichoptic (a) and dichoptic (b) visualisation (left and right eyes) of a 3D image, and the blend of colours for both objects (cube and pyramid) presented to the screen.}
    \label{fig:stereo-head}
}

\abstract{
    Adjusting transparency is a common method of mitigating occlusion but is often detrimental for understanding the relative depth relationships between objects as well as removes potentially important information from the occluding object. We propose using dichoptic opacity, a novel method for occlusion management that contrasts the transparency of occluders presented to each eye. This allows for better simultaneous understanding of both occluder and occluded. A user study highlights the technique's potential, showing strong user engagement and a clear preference for dichoptic opacity over traditional presentations. While it does not determine optimal transparency values, it reveals promising trends in both percentage and range that merit further investigation.
} 


\AddToShipoutPictureBG{%
  \put(575,75){\rotatebox{90}{\textsf{ This work has been submitted to the IEEE for possible publication. Copyright may be transferred without notice, after which this version may no longer be accessible.}}}
}

\begin{document}


\firstsection{Introduction}

\maketitle


Advancements in 3D data acquisition technologies, such as LiDAR, photogrammetry, and volumetric scanning, are enabling the capture of increasingly detailed and complex spatial datasets. As the fidelity and volume of 3D data grows, so too does the need for visualisation techniques capable of supporting effective exploration, interpretation, and decision-making. This growing richness in data exacerbates a long-standing challenge in 3D visualisation: occlusion. In 3D data visualisation, important structures are often partially or fully obscured by foreground elements, making it difficult for users to interpret spatial relationships or access critical internal features.
Traditional approaches to occlusion management, such as transparency adjustment, viewpoint manipulation, and cutaway views, offer only partial solutions. These techniques typically involve trade-offs between preserving contextual information and enhancing visibility of occluded content.

As binocular technology like head-mounted displays (HMDs) become more prevalent, the use of features not possible on traditional desktops have a wider potential for practical use. One of the most prominent uses of binocular technology is stereoscopy, where the presentation of two 3D environments at slightly different positions and orientations improves the depth perception. Stereoscopy has been seen to have benefits outside of improving depth perception with Alper et al \cite{alper_stereoscopic_2011} leveraging it to highlight points of interest within node-link diagrams. By presenting a target object to only one eye, Krekhov et al \cite{krekhov_deadeye_2019} present another use of binocular technology not possible on a traditional flat display, \textit{DeadEye}. Through a phenomenon known as binocular rivalry, where contrasting images presented to each eye fail to resolve, the target object became a pre-attentive visual cue. Both stereoscopy and \textit{DeadEye} are examples of dichoptic presentation, where two different images are presented to each eye, but at different extremes.

In this paper we introduce a novel application of dichoptic presentation for occlusion management in 3D stereoscopic visualisation. By purposefully introducing disparity into the opacity of occluders, dichoptic opacity, we believe it will elicit clearer visualisations and coherent interpretations of not only occluded points, but the occluder blocking it. Performed at the fragment shader level, rendering the opacity of an occluder higher in one eye than the other results in two similar images but with distinct focuses, occluder and occluded respectively. We conducted a user study focussing on how participants engage with dichoptic opacity by measuring their preferred opacity levels, independently varied between both eyes, for accessing multiple layers of a 3D surface while maintaining depth perception.

\section{Related Work}

\subsection{Stereoscopy}
Stereoscopy, an artificial depth cue, is achieved by leveraging binocular disparity, which is where two slightly offset views of the same environment are presented. Beyond the obvious improvement to depth perception gained through stereoscopy, it has also been shown to improve performance time during 3D data analysis tasks \cite{van_schooten_effect_2010} as well as Gerig et al \cite{gerig_missing_2018} finding evidence that "\textit{virtually recreated depth cues}" have a positive impact for reaching tasks. Additionally, stereoscopy has seen uses for highlighting regions within node-link diagrams, improving task performance \cite{alper_stereoscopic_2011}. Finally, in a comprehensive review of over 160 publications, \cite{mcintire_stereoscopic_2014} found that 60\% of experiments showed better performance with stereoscopic 3D displays than non-stereoscopic ones, proving most useful for object manipulation and finding/identifying objects and imagery.

\subsection{Occlusion Management \& Transparency}
Another important depth cue, but not one exclusive to binocular devices, is that of occlusion. Occlusion helps orient and position a target based on what is in front of, or behind, it. However, occlusion also hides important data that could be crucial for data analysis and exploration. Therefore, occlusion of 3D data is a topic that has been explored in detail with various attempts to mitigate it across traditional desktop 3D and immersive environments. An important aspect when dealing with occluded data is how it can be presented. In a review of over 50 occlusion visualisation techniques, Elmqvist et al \cite{elmqvist_taxonomy_2007} group them into five categories. One of these, virtual x-ray, refers to rendering occluding objects partly or fully transparent in order to allow a view of an occluded object behind it. Chen et al \cite{chen_global_2021} manage occlusion with transparency via an egocentric and scalable sphere that modifies the transparency of objects within its radius whereas Pietroszek et al \cite{pietroszek_tiltcasting_2015} presented an adjustable plane to hide objects in front of the plane.

The "\textit{Superman's X-ray vision}" problem defined by Livingston et al \cite{livingston_resolving_2003} refers to how removing occlusion via virtual x-rays makes the ordering of objects harder to discern as too much information prevents users from making sense of the relative depth relationships of objects. Elmqvist et al's image space algorithm for dynamic transparency \cite{elmqvist_employing_2007}, which renders only part of an object's surface transparent rather than the entire object, mitigates this problem by preserving information of the occluder with a window-like hole to view what is behind. Dichoptic opacity will also mitigate the problem, but with the added benefit of retaining the occluder as a cohesive whole due to more available information than would otherwise be possible in typical stereoscopic visualisations.

\subsection{Dichoptic Presentation}
Dichoptic presentation refers to when each eye is presented different images. When the two different images are similar enough to each other our brains are able to comprehend and fuse them into a single image we can understand. This is how stereoscopy works. But when the two images are too contrasting it prevents a single, cohesive image from being produced and instead our eyes swap focus between the two, something known as binocular rivalry. Krekhov et al \cite{krekhov_deadeye_2019} proposed a technique called \textit{Deadeye} where a pre-attentive visual cue was discovered as a result of binocular rivalry. By presenting to only one eye, they developed a visual cue that imposed no visual modifications to the target yet provided pop out. 


Despite a range of efforts to manage occlusion and utilise dichoptic presentation, existing approaches face limitations in either preserving spatial understanding or avoiding perceptual conflict. Transparency-based occlusion techniques risk degrading depth cues, while methods like DeadEye rely on perceptual disruption (binocular rivalry) rather than integration. 

Our approach introduces binocular disparity in opacity levels to manage occlusion without eliminating the occluder or compromising depth perception. By leveraging the human visual system’s ability to fuse asymmetric transparency information, our method aims to retain both the occluder's context and the occluded's visibility.

\section{Methodology}
To develop our approach, we designed a dichoptic visualisation technique and implemented it in the game engine Unity3D for experimental use on a stereoscopic display.

\subsection{Rationale}
Transparency-based methods need to carefully balance the perceptual prominence of both occluders and occluded; too much transparency can weaken depth cues, leading to things like "\textit{Superman's X-ray vision}" problem, while too little may hinder comprehension, leading to cognitive strain or misinterpretation. Dichoptic opacity is the divergent presentation of an occluder's transparency between the different displays of a binocular device, presenting the occluder more clearly in one eye and what is behind it more clearly in the other eye. Like how two images at slightly different orientations create a depth effect, stereoscopy, dichoptic opacity preserves and enables access to more information of occluder and occluded. With this additional information we can better preserve our spatial understanding, and retain the benefits of stereoscopy.

\subsection{Implementation}

We developed a version of dichoptic opacity in Unity 2022.3.5f1\footnote{https://unity.com/releases/editor/archive}.

An approach for the dichoptic presentation of opacity is to present overlapping stereoscopic 3D scenes, consisting of a left and right mesh for every visible object. Each version of the mesh is only presented to one eye using culling masks. This allows for the opacity to be effected individually between eyes by targeting the appropriate mesh. However this made the scene drastically more complex and time consuming to set up, with additional work needed to ensure both meshes for a given object remained consistent. Instead, dichoptic opacity was applied at the fragment shader level.

Using Unity's Universal Render Pipeline, we implemented dichoptic opacity through a custom shader pass: 
\begin{lstlisting}
	fixed4 frag(v2f i) : SV_Target {
		...
		int index = unity_StereoEyeIndex;
		float alpha = _LeftAlpha;
		
		if (index == 0) {
			index = _DisplayIndex;
		}
		if (_UseDichopticOpacity == 1) {
			alpha = _RightAlpha * index 
			+ _LeftAlpha * (1 - index);
		}
		return fixed4(finalColor, alpha);
	}
\end{lstlisting}

\textcolor{black}{The transparency of the object is handled through the fourth channel of the presented RGB colour, known as the alpha. Opaque objects have an alpha of 1 and completely transparent objects have an alpha of 0. When rendering a transparent object, depth checks are performed to see if it is being rendered in front of an existing object. If so, the alpha value is used to blend between the colour already at that point and the colour of the transparent object. With \textit{\_LeftAlpha} and \textit{\_RightAlpha} we change the alpha value independently between each eye}. These two properties are sequestered by a index based on either a camera's target display and determines the alpha value used during that object's rendering. \textcolor{black}{By directly modifying these alpha values, a participant can change the transparency of an object and receive an immediate response.}

\section{Evaluation}

We conducted a user study focussing on how participants engage with the dichoptic opacity technique by measuring their preferred opacity levels, independently varied between both eyes, for accessing multiple layers of a 3D surface while maintaining depth perception.
\subsection{Aim and Hypotheses}


This work lays the foundation for investigating the benefits of dichoptic opacity by first examining the variability between dichoptic and non-dichoptic opacity settings. \textcolor{black}{This is primarily an exploratory study to determine the potential of dichoptic opacity. We aim to identify whether opacity presented asynchronously offers advantages over traditional synchronous presentation by comparing subjective preferences for supporting spatial understanding and the interpretation of 3D structures when visualising partially transparent occluders. If there are any advantages, a secondary aim is to determine which type of threshold is preferred by users.}


We tested three hypotheses : There is a consensus regarding the optimal level of non-dichoptic opacity for an occluding layer in a 3D scene, to support the simultaneous interpretation of both front, occluder, and back, occluded, layers while maintaining spatial understanding (\textbf{H1}); when tasked with selecting the most accessible stereoscopic visualisation while preserving spatial understanding, participants will engage with dichoptic opacity by assigning significantly different transparency values to each eye (\textbf{H2}), there is a consistent pattern in the selected opacity values, both in terms of the difference between eyes and the average opacity across both eyes, indicating a consensus among participants on how to support the simultaneous interpretation of front and back layers (\textbf{H3}).

\subsection{Study Setup}
We designed and conducted a within-subjects experiment to test our hypotheses.
A pilot study was conducted with 2 participants to refine the task design and identify any technical or usability issues prior to the main study. \textcolor{black}{We found that the pilots failed to engage with the study, providing little input, indicating a lack of understanding with the tasks. As a result, for the full experiment we introduced pre-scripted instructions to remove ambiguity and inconsistency between participants' goals.}

\paragraph{Participants}
The main experiment was conducted with 11 participants (6 male, 5 female). All had normal or corrected to normal vision and 4 participants had previous experience with binocular technology in the form of VR headsets. \textcolor{black}{ Due to the exploratory nature of the study, and the few independent factors, this small pool of participants is suitable for gaining insights into the potential of dichoptic opacity before performing further complex studies that need more participants.}

\paragraph{Apparatus}
An executable program containing the task was ran on a Windows 10 computer with a NVIDIA RTX A4000 GPU. The computer running the experiment was linked via two 1920 x 1080 HDMI cables to an external display, a DRV-D1 Deep Reality Viewer  \footnote{https://www.visiontriteq.com/}; a stereo image presentation system that creates 3D images without the need for specialised eyewear or headsets. The DRV-D1 has two 1920 x 1080 resolution channels and a 16:9 aspect ratio 400 x 225 mm concave mirror as well as a 60Hz refresh rate.

\paragraph{Task Design}


We tested our hypothesis thanks to three tasks performed on a simplified 3D scene. The virtual scene consisted of two textured 3D meshes: an inner pyramid and an outer cube, with each mesh having distinct textures in both shape and colour to support visual differentiation. The pyramid retained its local position while both rotated at a consistent rate around an origin point.
The first task (\textbf{T1}) was performed under the non-dichoptic condition: a single opacity level was applied uniformly to both eyes. Participants, were initially encourage to explore the full opacity range, from minimum to maximum, to familiarise themselves with the transparency effect before selecting their preferred value. \textcolor{black}{Based on feedback from the pilot studies, the instructions to explore full range of values was emphasised with a written script.}

The second task (\textbf{T2}) was performed under the dichoptic condition: participants could adjust opacity independently for each eye. Participants were provided with their previously selected non-dichoptic opacity as a baseline, applied equally to both eyes. Similar to \textbf{T1}, participants were encouraged to explore the full opacity range \textcolor{black}{via a spoken script}, this time for both eyes independently, before selecting their preferred values.

For both tasks, opacity was controlled via keyboard input, with keys assigned for increasing and decreasing opacity. In \textbf{T2}, both eyes could be adjusted simultaneously or individually.

Participants were instructed to \textcolor{black}{change the transparency, via the alpha value,} of the occluding (outer) mesh until they could clearly perceive both meshes, aiming to maximise interpretability while maintaining spatial understanding.

In the final task (\textbf{T3}), participants were shown the results of both their dichoptic and non-dichoptic settings in a random order and asked to indicate which they preferred in terms of perceiving shape, detail, and overall spatial understanding.

\textcolor{black}{These tasks were each completed once in the scenario described.}



\begin{figure}
\centering
\includegraphics[alt={Two blue cubes with a yellow pyramid within each. The left cube is more transparent than the right so the left pyramid can be seen more clearly.},width=\linewidth]{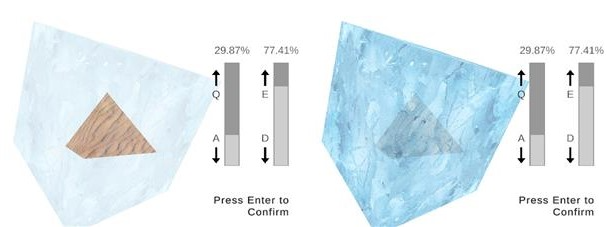}
\caption{Left: Image presented to left eye with more transparent cube so that the pyramid is more visible. Right: Image presented to right eye with more opaque cube so that the pyramid is harder to see}
\label{fig:t2-vis}
\end{figure}



\paragraph{Protocol}
Participants began with a short briefing on the study and then completed a consent form. After orienting in front of and adjusting the DRV-D1 device, they were provided with three tasks in succession, each of which was proceeded with on-screen written instructions. These instructions remained visible until the participant pressed a key (Enter) to confirm and begin the task. There were no time constraints on the tasks and breaks could be taken whenever needed. Participants completed the three tasks in order, \textbf{T1}, \textbf{T2}, and \textbf{T3}, advancing to the next task by pressing Enter. All additional instructions, beyond the written ones, were delivered orally and shared equally with all participants. These included encouragement to explore both the minimum and maximum opacity values before making their final selection. Tasks \textbf{T1}, \textbf{T2}, and \textbf{T3} lasted approximately 20 seconds, 30 seconds, and 5 seconds respectively \textcolor{black}{and were carried out once.} This concluded with a short debrief.

\subsection{Results Analysis}
\begin{figure}
    \centering
    \includegraphics[alt = {Graph of participants and their transparency values selected for both dichoptic and non-dichoptic, with a smaller difference at the top and larger at the bottom}, width=\linewidth]{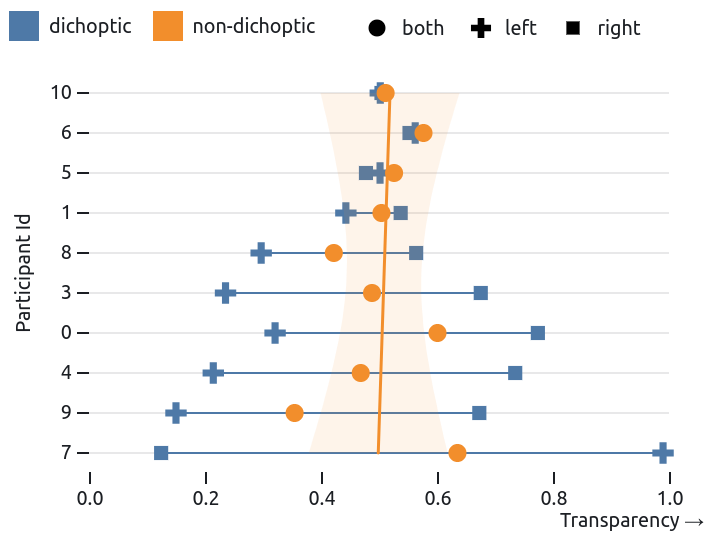}
    \caption{Graph showing opacity values selected by participants for the non-dichoptic condition (both eyes) and the dichoptic condition (left and right eyes), ordered by dichoptic opacity range. Data from P2 has been excluded due to outlier values.}
    \label{fig:exp-data}
\end{figure}

We analysed the participants' input and preferences to assess the potential of the dichoptic technique. The results for each task performed were stored and then generated into a JSON format at the completion of the experiment. All results were collected and formatted into a CSV file, which can be viewed as part of a GitHub repository \footnote{https://github.com/georgebellbell/Dichoptic-Opacity-Experiment-Data.git}. See Figure~\ref{fig:exp-data} for a visualisation of the data. 

One participant's results were excluded from the analysis due to a suspected misunderstanding of the task instructions which subsequently led to an erroneous result for the non-dichoptic opacity selected (0.91). This was identified as an outlier by both a high Z-Score (Z = 2.25) and being outside the IQR. This likely compromised the results and thus data for that participant was removed.

\paragraph{Non-dichoptic Condition}
For the non-dichoptic condition (\textbf{T1}), we measured the opacity level selected for the front layer of a stereoscopic 3D surface representation. The average opacity chosen across participants was M = 0.507 (SD = 0.084). The low standard deviation indicates strong agreement, thus supporting \textbf{H1} regarding participant consensus.

\paragraph{Dichoptic Condition}

The dichoptic selection consisted of two opacity values for the outer layer, one for each eye. We analysed the dichoptic midpoint, defined as the median of the left and right eye opacity values, and the dichoptic range, defined as the difference in opacity between the two eyes. The average dichoptic midpoint was M = 0.490 (SD = 0.0512). The average dichoptic range was M = 0.320 (SD = 0.289).  The high average for dichoptic range indicates strong user engagement, supporting \textbf{H2}, but its large standard deviation indicates no agreement, thus rejecting \textbf{H3}.

\paragraph{Participant Preference}
The participant preference was measured as a binary outcome: participants either preferred dichoptic opacity \textcolor{black}{for perceiving both object simultaneously} or they did not. The metric for this is the proportion of participants who preferred dichoptic opacity. This was calculated as the number of participants who preferred dichoptic opacity (7) divided by the number of participants (10). The result was 70\% showing a majority preference for dichoptic presentation.

\section{Discussion}

We reflected on the findings, considered limitations of our approach, and identified directions for future work.


The results of the user study indicate engagement and preference for dichoptic opacity, though further investigation is needed to validate these findings. There is a clear consensus between participants in regards to the non-dichoptic opacity used to distinguish both occluded and occluder in the scenario given in \textbf{T1}, supporting \textbf{H1}. The average opacity of 51\% can therefore be considered a reference point for comparison with the dichoptic condition. The non-dichoptic opacity represents a near equal mix of information between the occluding mesh and the occluded mesh within being presented simultaneously. Regarding the dichoptic condition, we observe strong engagement, as indicated by the high average range difference between the two eyes, supporting \textbf{H2}. However, this engagement is not consistent across participants, preventing us from concluding a clear preference for specific opacity values, discarding \textbf{H3}. The average opacity across both eyes is approximately 49\%, close to the 51\% observed in the non-dichoptic condition. This suggests that, although participants engaged with interocular disparity, they tended to maintain an overall opacity level similar to that in the non-dichoptic condition. 

Majority of the participants preferred dichoptic opacity with only 3 of the 10 preferring non-dichoptic, showing a willingness to use dichoptic opacity. \textcolor{black}{With the 7 participants who preferred dichoptic opacity providing a dichoptic range above the lower quartile, 0.0418, it indicates that the difference contributes significantly towards user preference. In support of this,} of the 3 participants who preferred non-dichoptic, 2 had a dichoptic range below the lower quartile. \textcolor{black}{However,} this may have resulted in distinguishing between options in \textbf{T3} being harder and thus less reliably indicates preference for the technique. In contrast, one participant (P7) selected an unusually large opacity range, which we believe to be detrimental to stereoscopic perception. However, it remains unclear whether this setting was still sufficient for the participant to access depth, or whether they simply disregarded the instruction to preserve spatial understanding. This highlights the need to consider the spatial understanding more carefully in future work.


Future work will need to explore different conditions and scenarios when presenting dichoptic opacity. This includes comparing different levels of contrast between the opacity values presented between each eye, increasing at regular intervals to determine optimal opacity values for perceiving shape and detail, as well as preserving spatial understanding of occluder and occluded. \textcolor{black}{Finally it is acknowledged that for this study two patterns of different colours are used. There is a possibility that the introduction of other colours, be they homogeneous or another texture, may have had an impact of the resulting optimal opacity value. However we believe that it would have an insignificant impact on this preliminary study and therefore has not been considered. This could be looked at more carefully in future work.}

%



\section{Conclusion}


In summary, we propose the use of dichoptic opacity as a novel approach to occlusion management for rendering 3D surface data in stereoscopic environments. Our technique contributes to research pertaining to both stereoscopic visualisation and occlusion management by contrasting occluder opacity between each eye to allow simultaneous interpretation of more information for occluder and occluded while preserving spatial understanding.

To test the potential of dichoptic opacity we conducted a user study looking at how participants engaged with dichoptic opacity. Across three tasks, participants were asked to select and then compare preferred opacity values for dichoptic and non-dichoptic opacity. Our results showed that not only did participants engage with dichoptic opacity, selecting significantly different opacity values between each eye, but majority preferred it to traditional opacity when asked to perceive two objects simultaneously.

The results from this initial experiment presented a broad range of values for dichoptic opacity and therefore it will be a primary focus of future work to determine what values are most optimal for the technique. Additionally, work will be needed to evaluate the performance of dichoptic opacity when compared to non-dichoptic opacity as well as any correlations between the two. With the expressed interest for dichoptic opacity found during this experiment, it will be important for future work to address potential integration of the technique with existing occlusion management techniques.

By continuing to use binocular technology in ways outside of standard stereoscopic 3D presentations, it can further our perception and interpretation of complex data in ways otherwise not possible on traditional displays.
\newpage
\bibliographystyle{abbrv-doi}

\bibliography{template}
\end{document}